\documentclass{PoS}

\usepackage{citesort}
\usepackage{graphicx}
\usepackage{dcolumn}
\usepackage{bm}
\usepackage{xspace}
\usepackage{epsfig}

\usepackage{amsmath}
\usepackage{amssymb}

\newcommand{\dzero}     {D\O\xspace}
\newcommand{\ttbar}     {\ensuremath{t\overline{t}}\xspace}
\newcommand{\ppbar}     {\ensuremath{p\overline{p}}\xspace}

\newcommand{\met}       {\mbox{$\not\!\!E_T$}\xspace}

\newcommand{\wplus}     {$W$+jets\xspace}

\newcommand{\kapas}     {MC-to-data correction factors\xspace}

\newcommand{\muplus}    {$\mu$+jets\xspace}
\newcommand{\eplus}     {$e+$jets\xspace}
\newcommand{\lplus}     {$\ell+$jets\xspace}

\newcommand{\sigmaB}    {$\sigma_{X} \times B(X\rightarrow \ttbar)$}

\title{Search for a new Resonance decaying into Top-Antitop at Tevatron}

\ShortTitle{Search for a new Resonance decaying into Top-Antitop at Tevatron}

\author{\speaker{Christian Schwanenberger}\thanks{On behalf of the CDF
        and \dzero Collaborations.}\\
        Physikalisches Institut der Universit{\"a}t Bonn, Germany\\
        E-mail: \email{schwanen@fnal.gov}}

\abstract{
In this report a new search for a narrow-width heavy resonance decaying into top quark pairs ($X\rightarrow\ttbar$) 
in \ppbar collisions at $\sqrt{s}=1.96$~TeV has been performed using data collected by 
the {\dzero} detector at the Fermilab Tevatron collider. 
The analysis considers \ttbar candidate events in the lepton+jets 
channel using a lifetime tag to identify $b$-jets and the \ttbar invariant mass distribution 
to search for evidence of resonant production. The analyzed dataset corresponds to an
integrated luminosity of approximately $370$ pb$^{-1}$.
Since no evidence for a \ttbar resonance $X$ is found, upper limits on {\sigmaB} 
for different hypothesized resonance masses using a Bayesian approach
are set.
Within a topcolor-assisted technicolor model, the existence of a leptophobic $Z'$ boson with 
$M_{Z'} < 680$~GeV and width $\Gamma_{Z'} = 0.012 M_{Z'}$ can be excluded at $95\%$ C.L..
}

\FullConference{International Europhysics Conference on High Energy Physics\\
		 July 21st - 27th 2005\\
		 Lisboa, Portugal}

\begin{document}

\section{\label{sec:intro}Introduction}
The top quark has by far the largest mass of all known elementary particles. This
suggests that the top quark may play a special role in the dynamics
of electroweak symmetry breaking. One of the various models incorporating this possibility
is topcolor~\cite{topc1}, where the large top quark mass can be
generated through a dynamical {\ttbar} condensate, $X$, which is
formed by a new strong gauge force preferentially coupled to the third
generation of fermions. 
In one particular model, topcolor-assisted technicolor~\cite{topc2}, $X$ couples
weakly and symmetrically to the first and second generations and
strongly to the third generation of quarks, and has no couplings to
leptons, resulting in a predicted cross section for {\ttbar} production
larger than the standard model (SM) prediction.

In this presentation a new model-independent search for a narrow-width
heavy resonance $X$ decaying into  
{\ttbar} is presented. The analyzed dataset corresponds to an
integrated luminosity of $366\pm 24$\,pb$^{-1}$ in the \eplus 
channel and $363\pm 24$ pb$^{-1}$ in the \muplus channel, collected
between August 2002 and August 2004. 

\section{\label{sec:search}Search for \ttbar production via new resonances}

%
In the framework of the SM, the
top quark decays into a $W$ boson and $b$ quark in nearly 100\% of the
cases.
The \ttbar event signature is fully determined by the $W$ boson decay
modes. 
In this analysis only the lepton+jets ($\ell$+jets, where $\ell = e
$~or~$\mu$) final state, which results from the 
leptonic decay of one of the $W$ bosons and the hadronic decay of the
other, is considered. The investigated event 
signature is one isolated electron ($|\eta|<1.1$) or muon
($|\eta|<2.0$) with high transverse
momentum $p_T>20$~GeV,  
large transverse energy imbalance $\met>20$~GeV due to the undetected
neutrino, and at least four jets (defined using a cone algorithm with radius 
$\Delta{\cal R}=0.5$) with $p_T>15$~GeV and rapidity $|y|<2.5$,  
two of which result from the hadronization of the $b$ quarks. Further
details of the event selection and of the following can be found
in~\cite{confnote}. 

The signal-to-background ratio is improved by identifying $b$-jets
using a lifetime based $b$-tagging algorithm. 
After $b$-tagging, the dominant physics background for a resonance signal is 
non-resonant SM {\ttbar} production. Smaller contributions arise from
the direct production of $W$ bosons in
association with four or more jets (\wplus), as well as instrumental
background originating from 
multijet processes with jets faking isolated leptons. 

\begin{figure}
\vspace{-0.35cm}
\begin{center}
\begin{tabular}{cc}
\epsfig{file=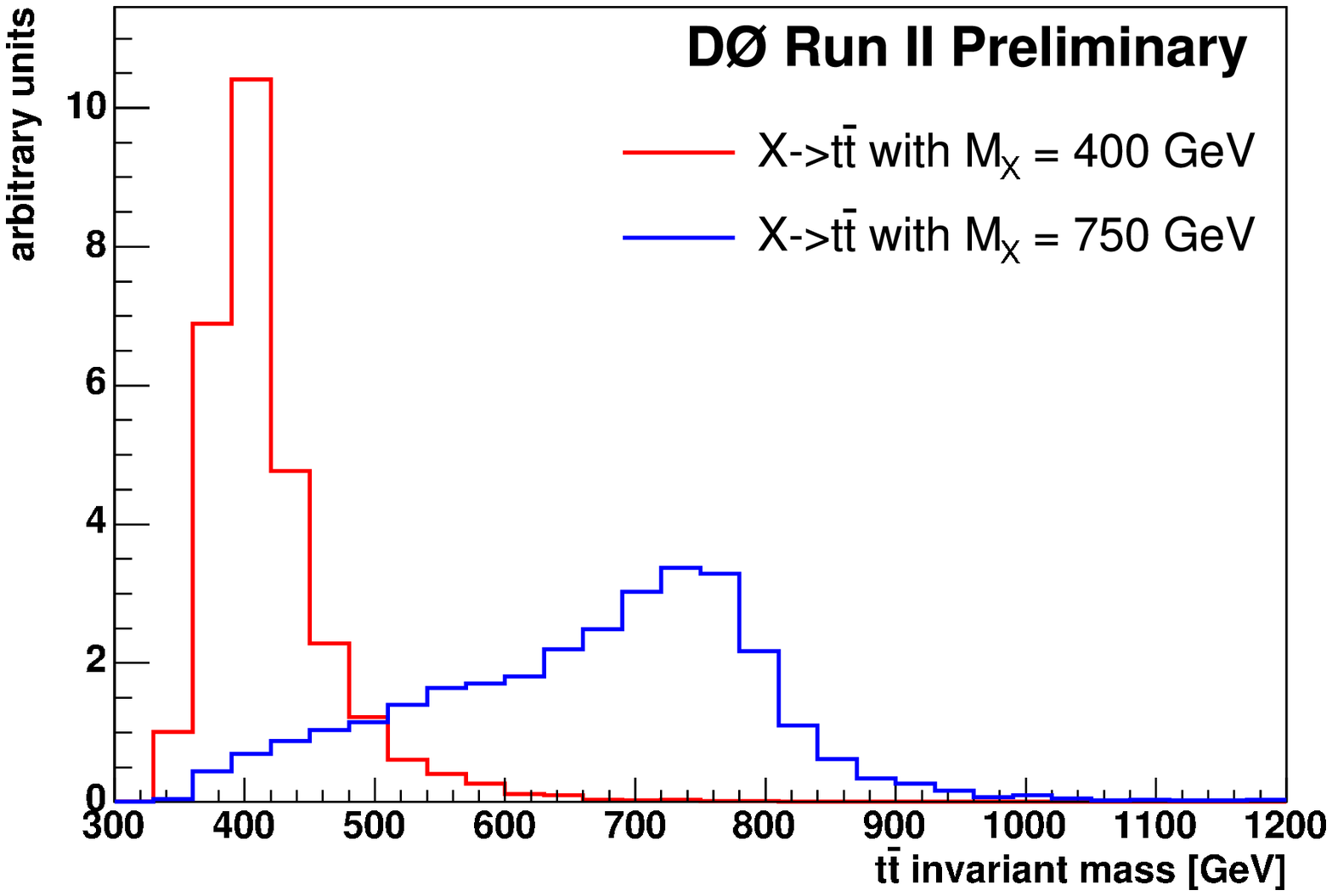,width=7.5cm} & 
\epsfig{file=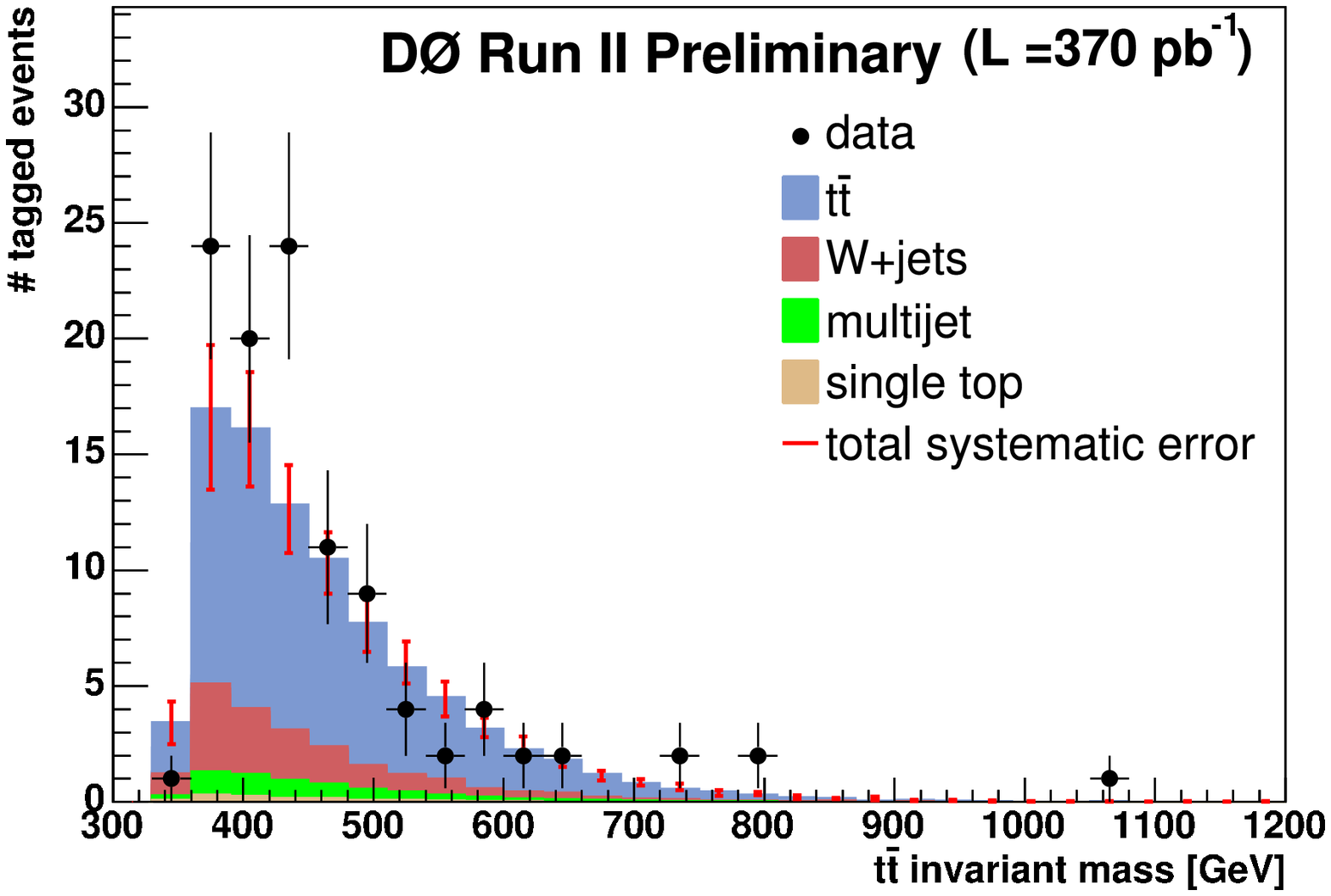,width=7.5cm} \\
\end{tabular}
\caption{\label{fig:invmass}Comparison of the expected \ttbar invariant
  mass distribution for a  
narrow-width resonance with mass $M_X=400$~GeV and $M_X=750$~GeV
  (left). The resulting \ttbar invariant mass distribution for the
  combined \lplus channels (right). The error bars drawn on top of the
  SM background indicate the total systematic uncertainty, which has 
significant bin-to-bin correlations.}
\end{center}
\end{figure}

%
%
The search for
resonant production is performed 
by examining the reconstructed \ttbar invariant mass distribution
resulting from a constrained kinematic fit to the \ttbar
hypothesis. The used fit is similar to the one used for the
measurement of the top quark mass in 
Run I~\cite{kinfit}. The constraints to the fit are that 
two jets and the lepton+\met must each form the invariant
mass of the $W$ boson and that the masses of the two reconstructed top quarks have to be equal, and are set to $175$ GeV.
Only the four highest $p_T$ jets are considered in the kinematic fit.
From the resulting twelve possible jet-parton assignments, the one with the lowest 
$\chi^2$ is chosen. This is found to give the correct solution in about $65\%$ of the \ttbar events.
The expected \ttbar invariant mass distribution for resonance 
masses of $400$~GeV and $750$~GeV are illustrated in
Figure~\ref{fig:invmass} (left).

%
%
The systematic uncertainties of the analysis rely on the prediction of
the overall normalization as 
well as the shape of the reconstructed \ttbar  
invariant mass distribution for both signal and the different
backgrounds. The systematic uncertainties can 
be classified as those affecting only normalization and those
affecting both normalization and shape of the 
\ttbar invariant mass distribution.

The systematic uncertainties affecting only the normalization include
e.g. the experimental uncertainties on the  
\kapas, the theoretical uncertainty on the SM prediction for
$\sigma_{\ttbar}$, 
$\sigma_{single top}$ and the uncertainty on the integrated 
luminosity.
The systematic uncertainties affecting the shape of the \ttbar
invariant mass distribution in addition  
to the normalization have been studied both on signal and background samples.
These include e.g. uncertainties on the jet energy calibration, jet
reconstruction efficiency,  $b$-tagging 
parameterizations for $b$, $c$ and light jets and the limited top quark
mass accuracy as it enters the kinematic fit.

The systematic uncertainties associated with the estimation of
the fractions for the different flavor components of the \wplus
background have been taken into account as well as 
the uncertainty associated with the modeling of the SM \ttbar
background, in particular, gluon radiation effects. 

\section{\label{result}Result}

In the final selection 57 events remain in the \eplus channel and 51
events in the \muplus channel. 
Figure~\ref{fig:invmass} (right) shows the \ttbar invariant mass for
the combined $\ell$+jets channels for the selected events in 
data and for the SM background predictions. 

Assuming there is no resonance signal, a Bayesian approach is used to
calculate $95\%$ C.L. upper limits on 
{\sigmaB} for different hypothesized values for $M_X$.
A Poisson distribution is assumed for the number of observed events in
each bin, as well as  
flat prior probabilities for the signal cross section. Systematic
uncertainties on the signal acceptance 
and background yields are implemented via a convolution procedure of a
multivariate Gaussian distribution 
implementing a full covariance matrix including correlations.

\begin{figure}
\vspace{-0.35cm}
\begin{center}
\begin{tabular}{c}
\epsfig{file=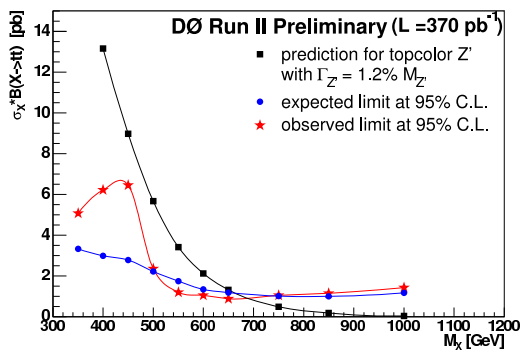,width=7.5cm} \\
\end{tabular}
\caption{\label{fig:limsum}Expected and observed 95$\%$ C.L. upper limits on {\sigmaB} compared with 
the predicted topcolor-assisted technicolor cross section for a $Z'$ boson with a width of 
$\Gamma_{Z'} = 0.012 M_{Z'}$ as a function of resonance mass $M_{X}$.}
\end{center}
\end{figure}

The expected and observed $95\%$~C.L. upper limits on {\sigmaB} as a function of $M_X$ are displayed in Fig.~\ref{fig:limsum}. This figure also includes the
predicted {\sigmaB} for a leptophobic Z' boson with $\Gamma_{Z'} =
0.012 M_{Z'}$ which, combined with 
the experimental limits, allows to exclude $M_{Z'} < 680$~GeV at
$95\%$ C.L.. 

Similar searches were performed at $\sqrt{s}=1.8$ TeV by the CDF and
\dzero collaborations during 
Run I of the Tevatron collider analyzing a data set of
$106$\,pb$^{-1}$ (CDF) and $130$\,pb$^{-1}$ (D\O), respectively. No
evidence for a \ttbar resonance was found. 
The resulting limits on {\sigmaB}, where $\sigma_{X}$ is the resonance production cross section,
were used to exclude a leptophobic Z' boson with $\Gamma_{Z'} = 0.012 M_{Z'}$. The excluded mass
regions at $95\%$ C.L. are, respectively, $M_{Z'}<480$~GeV~\cite{RunIttrescdf}
and $M_{Z'}<560$~GeV~\cite{RunIttresdzero}.
Thus the new Run II measurement presented here extends 
the \dzero Run I exclusion on $M_{Z'}$~\cite{RunIttresdzero} by 120 GeV.

\section{\label{conc}Conclusion}
A search for a narrow width resonance in the \lplus final states has
been performed using data  
corresponding to an integrated luminosity of about $370$~pb$^{-1}$,
collected with the \dzero detector 
during Run II of the Tevatron collider. By analyzing the reconstructed
\ttbar invariant mass 
distribution and using a Bayesian method, model independent upper
limits on {\sigmaB} have 
been obtained for different hypothesized masses of a narrow-width
heavy resonance decaying into \ttbar. 
Within a topcolor-assisted technicolor model, the existence of a leptophobic $Z'$ boson with $M_{Z'} < 680$~GeV and width $\Gamma_{Z'} = 0.012 M_{Z'}$ can be excluded at $95\%$ C.L..

\end{document}